\documentclass[prd,aps,12pt]{article}

\usepackage{epsf}
\usepackage{amsmath, amssymb}

\begin{document}
  
\begin{titlepage}

\def\thefootnote{\fnsymbol{footnote}}

\begin{center}

\hfill TU-831 \\
\hfill UT-HET-017  \\
\hfill November, 2008

\vspace{0.5cm}
{\Large\bf Synchrotron Radiation from the Galactic Center \\
in Decaying Dark Matter Scenario}

\vspace{1cm}
{\large Koji Ishiwata}$^{\it (a)}$\footnote{E-mail: 
ishiwata@tuhep.phys.tohoku.ac.jp},
{\large Shigeki Matsumoto}$^{\it (b)}$\footnote{E-mail: 
smatsu@sci.u-toyama.ac.jp},
{\large Takeo Moroi}$^{\it (a)}$\footnote{E-mail: 
moroi@tuhep.phys.tohoku.ac.jp}

\vspace{1cm}

{\it $^{(a)}${Department of Physics, Tohoku University,
    Sendai 980-8578, Japan}}

\vspace{0.5cm}

{\it $^{(b)}${Department of Physics, University of Toyama, 
    Toyama 930-8555, Japan}}

\vspace{1cm}
\abstract{ 

  We discuss the synchrotron radiation flux from the Galactic center
  in unstable dark matter scenario.  Motivated by the anomalous excess
  of the positron fraction recently reported by the PAMELA
  collaboration, we consider the case that the dark matter particle is
  unstable (and long-lived), and that energetic electron and positron
  are produced by the decay of dark matter.  Then, the emitted $e^\pm$
  becomes the source of the synchrotron radiation.  We calculate the
  synchrotron radiation flux for models of decaying dark matter, which
  can explain the PAMELA positron excess.  Taking the lifetime of the
  dark matter of $O(10^{26}\ {\rm sec})$, which is the suggested
  value to explain the PAMELA anomaly, the synchrotron radiation flux
  is found to be $O(1\ {\rm kJy/str})$ or smaller, depending on the
  particle-physics and cosmological parameters.

}

\end{center}
\end{titlepage}

\renewcommand{\theequation}{\thesection.\arabic{equation}}
\renewcommand{\thepage}{\arabic{page}}
\setcounter{page}{1}
\renewcommand{\thefootnote}{\#\arabic{footnote}}
\setcounter{footnote}{0}

\section{Introduction}
\label{sec:intro}
\setcounter{equation}{0}

Nowadays there is no doubt about the existence of dark matter in our
universe. However, nobody knows what dark matter is. The answer for
the question will give a great impact on various fields of physics
including astrophysics, cosmology and particle physics. For
astrophysics and cosmology, dark matter is the dominant source of
gravitation, and it governs the structure formation of our
universe. For particle physics, we cannot find the candidate for dark
matter in the standard model, so that dark matter is now regarded
as the first evidence of new particles beyond the standard model.

Experimentally, many attempts have been made to understand the
property of dark matter. Among them, the PAMELA collaboration has
recently reported quite an interesting result; the anomalous excess of
positrons with the energy of 20 $-$ 100 GeV in the cosmic ray
\cite{Adriani:2008zr}. The excess is hardly understood by the
conventional mechanism \cite{Moskalenko:1997gh, Baltz:1998xv}, namely
the secondary production of positrons due to the collision between
primary protons and interstellar medium in our Galaxy.

Theoretically, a lot of scenarios for dark matter have been proposed
to explain the anomalous excess of positrons. These scenarios are
categorized into two types; positron productions from the annihilation
of dark matter \cite{Cirelli:2008pk, Cholis:2008qq, Feldman:2008xs,
  Fox:2008kb,Bergstrom:2008gr,Barger:2008su,Nelson:2008hj,Harnik:2008uu}
and from the decay of dark matter \cite{Yin:2008bs, Ishiwata:2008cv,
  Chen:2008md, Hamaguchi:2008rv, Ponton:2008zv, Ibarra:2008jk}. In
most of the scenarios in the former case, people have faced to the
difficulty to explain the excess; a huge enhancement factor, called
boost factor, to the production rate of positrons is needed, where the
factor is from the inhomogeneity of the dark matter distribution in
the vicinity of the solar system.\footnote
{There also be some exceptions, for instance, a huge annihilation cross
  section can be obtained if the Sommerfelt enhancement to the cross
  section is considered \cite{Hisano:2002fk,
  Hisano:2003ec,ArkaniHamed:2008qn}. A large boost factor is not
  mandatory in those scenarios}
On the contrary, in the latter case, the positron flux to explain the
excess can be naturally obtained with the lifetime of dark matter much
longer than the age of the universe. One might worry about the
fine-tuning problem to have a long lifetime of dark matter, however,
there are some scenarios to realize such a long lifetime, as we
discuss in the following.

Synchrotron radiation from the annihilation or decay of dark matter
may give constrains directly to scenarios aiming to the explanation of
the PAMELA positron excess. Since dark matter annihilates or decays
into energetic positron (as well as electron) under the circumstance
of magnetic fields in our galaxy, synchrotron radiation is inevitably
induced. Importantly, the WMAP collaboration has observed the
radiation in the whole sky, so that the observation gives constraints
on the scenarios of the electron and positron production due to the
annihilation or decay of dark matter in the Galactic halo.  The
synchrotron radiation flux for the case of dark matter annihilation
has been considered in past works \cite{Hooper:2007kb, Hooper:2008zg}.
The purpose of this article is to study the synchrotron radiation from
the decay of dark matter, and discuss implication to the scenarios of
decaying dark matter which account for the PAMELA result.

The organization of this paper is as follows.  In the next section, we
explain the cosmological scenario we consider.  The formalism to
calculate the synchrotron flux from the decay of dark matter is given
in Section \ref{sec:synchotron}.  Compared to the previous works,
several improvements are realized.  Then, we numerically calculate the
flux for several dark matter models; the results are shown in Section
\ref{sec:results}.  The summary of our study is given in Section
\ref{sec:summary}.

\section{The Scenario}
\label{sec:scenario}
\setcounter{equation}{0}

In this section, we summarize the cosmological scenario we consider.
As we mentioned in the introduction, taking the anomalous excess of
the positron fraction measured by the PAMELA \cite{Adriani:2008zr}
seriously, we consider unstable dark matter scenario.  Indeed, if the
decay of dark matter produces energetic positron, it becomes the
source of the cosmic ray positron, and may explain the excess of the
positron fraction.

In the following, we consider the case that the dark matter particle
(which we denote $X$) is unstable, and that its decay produces
energetic positron (as well as electron).  We also assume that the
energetic positron produced by the decay is the origin of the excess
of the positron fraction observed by the PAMELA.  So far, many
possibilities of such an unstable dark matter has been discussed
\cite{Yin:2008bs, Ishiwata:2008cv, Chen:2008md, Hamaguchi:2008rv,
  Ponton:2008zv, Ibarra:2008jk}.  The energy spectrum of the produced
$e^\pm$ depends on the model; in some cases, only the monochromatic
electron and positron are produced by the decay and, in other cases,
the energy distribution of the $e^\pm$ is non-monochromatic.

The positron flux from the dark matter decay is inversely proportional
to the lifetime of dark matter $\tau_{\rm DM}$.  The best-fit value of
$\tau_{\rm DM}$ to explain the PAMELA anomaly depends on the energy
spectrum of the positron from the dark matter decay.  For example, if
the dark matter mostly decays as $X\rightarrow e^+e^-$, the PAMELA
anomaly is well explained when \cite{Ishiwata:2008cv}
\begin{eqnarray}
  \tau_{\rm DM} \simeq 2.2\times 10^{27}\ {\rm sec}
  \times \left( \frac{m_{\rm DM}}{100\ {\rm GeV}} \right)^{-1}.
  \label{tau_ee}
\end{eqnarray}
For example, if the sneutrino lightest superparticle (LSP) in the
supersymmetric model is dark matter and also if it decays via an
$R$-parity breaking superpotential interaction, this is the
case.\footnote
{Even if the sneutrino is dark matter, 
the direct-detection constraint can be avoided.  See \cite
{Hall:1997ah, Asaka:2005cn}.}
On the contrary, if dark matter is the unstable gravitino which
dominantly decays into $l_i^\pm W^\mp$ final state (with $l_i^\pm$
being charged lepton in $i$-th generation)
\cite{Ibarra:2008qg,Ishiwata:2008cu}, $e^\pm$ is also produced by the
decay of $W^\pm$.  In such a case, the best-fit lifetime becomes
\begin{eqnarray}
  \tau_{\rm DM} \simeq 8.5\times 10^{26}\ {\rm sec}
  \times \left( \frac{m_{\rm DM}}{100\ {\rm GeV}} \right)^{-1}.
  \label{tau_gravitino}
\end{eqnarray}
In any case, in order to explain the PAMELA anomaly, the lifetime of
$O(10^{26-27}\ {\rm sec})$ is needed (when $m_{\rm DM} \sim O(100\
{\rm GeV})$).  In the following, we concentrate on the case that
$\tau_{\rm DM}$ is of this order.  In addition, as we will discuss,
taking these values of the lifetime and $m_{\rm DM}\sim 1\ {\rm TeV}$,
the $(e^-+e^+)$ flux becomes consistent with those reported by the
ATIC \cite{ATIC} and the PPB-BETS \cite{Torii:2008xu} experiments,
which have observed excess of the flux at the energy of a few hundred
GeV.

Once the electron and positron are emitted by the decay of dark matter
in our Galaxy, they cause the synchrotron radiation and the emitted
radiation may be observed.  In the following, we calculate the flux of
the synchrotron radiation from the central region of our Galaxy.

\section{Synchrotron Radiation Flux}
\label{sec:synchotron}
\setcounter{equation}{0}

In this section, we formulate the calculation of the radio flux from
synchrotron radiation generated by high energy electron (and
positron).  First, we discuss the propagation of the electron and
positron produced by the decay in the Galaxy.  Then, we summarize the
formula for the calculation of the synchrotron radiation.

\subsection{Propagation model for the electron and positron}

The synchrotron radiation flux depends on the energy spectra of the
electron and positron (at each point of the Galaxy) produced by the
dark matter decay.  We first discuss the propagation of the electron
and positron in the Galaxy to understand how the spectra are
determined.  For the electron and positron from the decay of dark
matter, their spectra are expected to be the same, so we consider the
electron spectrum.

When the energetic electron is emitted, its trajectory is twisted in
the so-called diffusion zone where the magnetic field is
non-negligible.  Because the magnetic field is entangled, the motion
of the electron is expected to be well approximated by a random walk.
Then, the propagation of the electron is described by the following
differential equation:
\begin{eqnarray}
  K(E) \nabla^2 f_{e}(E,\vec{x})
  + \frac{\partial}{\partial E}
  \left[ b(E,\vec{x}) f_{e}(E,\vec{x}) \right]
  + Q(E,\vec{x}) = 0,
  \label{DiffEq}
\end{eqnarray}
where $f_{e}$ is the electron spectrum (i.e., number density of the
electron per unit energy), $K(E)$ is the diffusion coefficient,
$b(E,\vec{x})$ is the energy loss rate, and $Q(E,\vec{x})$ is the
electron source term.

\begin{table}[t]
  \begin{center}
    \begin{tabular}{lcccc}
      \hline \hline
      Model & $\delta$ & $K_0\ ({\rm kpc^2/Myr})$ & $L ({\rm kpc})$ 
      & $R ({\rm kpc})$ \\ \hline
      M1 & $0.46$ & $0.0765$ & $15$ & $20$ \\
      MED~~ & $0.70$ & $0.0112$ & $4$ & $20$ \\ 
      M2 & $0.55$ & $0.0060$ & $1$ & $20$ \\
      \hline \hline
    \end{tabular}
    \caption{Parameters for the propagation models of electron, which
      will be used in our numerical calculations.}
    \label{table:propagatiomparam}
  \end{center}
\end{table}

Diffusion of injected electron is caused by the entangled magnetic
field in the Galaxy.  The function $K(E)$ can be determined so that
the cosmic-ray Boron to Carbon ratio and sub-Fe to Fe ratio are
reproduced.  In our analysis, we adopt the propagation model given in
\cite{Delahaye:2007fr}, where the shape of the diffusion zone is
approximated by a cylinder (with the radius $R$ and the half-height
$L$) and the function $K(E)$ is parametrized as
\begin{eqnarray*}
  K (E) = K_0
  \left(\frac{E}{1\ {\rm GeV}}\right)^\delta.
\end{eqnarray*}
Parameters of three representative models, MED, M1, and M2 models, are
shown in Table \ref{table:propagatiomparam}.  The MED model is the
best-fit to the Boron-to-Carbon ratio analysis, while the maximal and
minimal positron fractions for $E\gtrsim 10\ {\rm GeV}$ are expected
to be estimated with M1 and M2 models, respectively.

The energy loss of electron is via synchrotron radiation under the
magnetic field and inverse Compton scatterings with cosmic microwave
background (CMB) and infrared gamma ray from stars.  Then,
$b(E,\vec{x})$ is given by the sum of the synchrotron energy loss rate
$P_{\rm synch}$ and the inverse Compton energy loss rate $P_{\rm IC}$:
\begin{eqnarray}
  b(E,\vec{x}) 
  = 
  P_{\rm synch}+P_{\rm IC}
  \equiv
  P_{\rm synch}
  \left[ 
    1+r_{\rm IC/synch}(\vec{x}) 
  \right].
  \label{fn_b}
\end{eqnarray}
For simplicity, we neglect the position dependence of the magnetic
flux density $B$ in the diffusion zone; then, $P_{\rm synch}$ becomes
uniform in the diffusion zone.  Since the magnetic flux in the Galaxy
is not well understood, we take several values of $B$ to see how the
synchrotron radiation flux depends on the magnetic field in the
following numerical analysis.  The synchrotron energy loss rate
$P_{\rm synch}$ is given by \cite{Rybicki, Longair}
\begin{eqnarray}
  P_{\rm synch}
  = 
  \frac{e^4E^2B^2}{6\pi \epsilon_0 m_e^4c^5}
  \simeq
  3.4 \times 10^{-17} \ {\rm GeV sec}^{-1} \times
  \left(\frac{E}{1\ {\rm GeV}}\right)^2
  \left(\frac{B}{3 \ \mu{\rm G}}\right)^2,
  \label{eq:Psynch}
\end{eqnarray}
where $e$ is the electron electric charge, $\epsilon_0$ is the
permittivity of free space, $m_e$ is the electron mass, and $c$ is the
speed of light.  In addition, $r_{\rm IC/synch}$ is expressed as
\begin{eqnarray}
%
  r_{\rm IC/synch} (\vec{x})
  = 
  \frac{2}{3}
  \frac{U_{\rm rad}(\vec{x})}{(B^2/2\mu_0)}
  \simeq
  2.7 \times 
  \left( \frac{U_{\rm rad}}{0.9\ {\rm eV cm}^{-3}} \right)
  \left(\frac{B}{3 \ \mu{\rm G}}\right)^{-2},
  \label{eq:ratio}
\end{eqnarray}
where $U_{\rm rad}$ is the radiation density, and $\mu_0$ is the
permeability of free space.  In order to take into account the
position dependence of $U_{\rm rad}$, we use the results given in
\cite{Strong:1998fr} where the radiation density on the Galactic disc
(i.e., $U_{\rm rad}(r_\parallel,z=0)$) and that on the surface of
$r_\parallel =4\ {\rm kpc}$ (i.e., $U_{\rm rad}(r_\parallel =4\ {\rm
  kpc},z)$) are given, where $z$ and $r_\parallel$ are the height and
the radial distance in the cylindrical polar coordinate.  We assume
the following formula for the radiation density (unless otherwise
mentioned):
\begin{eqnarray}
  U_{\rm rad}(r_\parallel,z) = 
  \frac{U_\star(4\ {\rm kpc},z)}{U_\star(4\ {\rm kpc},0)} 
  U_\star(r_\parallel,0)
  + U_{\rm CMB},
\end{eqnarray}
with $U_{\rm CMB}\simeq 0.3\ {\rm eV cm}^{-3}$ being the CMB
contribution to the radiation density.  In addition, $U_\star\equiv
U_{\rm rad}-U_{\rm CMB}$ is the radiation density from stars, which is
determined from the results given in \cite{Strong:1998fr}.  Then, one
can see that $U_{\rm rad}$ varies as 0.3$-$10 eVcm$^{-3}$ from the rim
to the center of the Galaxy.  Notice that, in many previous works,
$U_{\rm rad}$ is approximated to be a constant.  However, the position
dependence of $U_{\rm rad}$ gives a significant effect on the
resultant synchrotron radiation flux because the inverse Compton
scattering may become the dominant energy-loss process.

The electron source term is given by electron injection rate and dark
matter distribution in the Milky Way Galaxy halo as
\begin{eqnarray}
  Q(E,\vec{x})= \frac{1}{\tau_{\rm DM}}
  \frac{\rho_{\rm DM}(\vec{x})}{m_{\rm DM}}
  \frac{dN_{e}}{dE},
  \label{sourceterm}
\end{eqnarray}
where $\rho_{\rm DM}$ is energy density of dark matter and $dN_{e}/dE$
is energy distribution of electron from the decay of single $X$.  For
the dark matter distribution $\rho_{\rm DM}$, we use the isothermal
halo density profile:
\begin{eqnarray}
  \rho_{\rm halo}(r) 
  =
  \rho_\odot \frac{r_{\rm core}^2 + r_\odot^2}{r_{\rm core}^2+r^2},
  \label{eq:isothermal}
\end{eqnarray}
where $\rho_\odot\simeq 0.43\ {\rm GeV/cm^3}$ is the local halo
density, $r_{\rm core}\simeq 2.8\ {\rm kpc}$ is the core radius,
$r_\odot\simeq 8.5\ {\rm kpc}$ is the distance between the Galactic
center and the solar system, and $r$ is the distance from the Galactic
center.

Before solving the diffusion equation \eqref{DiffEq}, it is
instructive to estimate the typical propagation length of the electron
per time scale of the energy loss, which we denote $\langle
l(E)\rangle$.  For the electron with energy $E$, the time scale of the
energy loss is $\sim E/b(E)$, and hence we obtain $\langle
l(E)\rangle\sim\sqrt{KE/b}$.  For the propagation model given in Table
\ref{table:propagatiomparam}, $\langle l(E)\rangle$ becomes a few kpc
for $E\sim 1\ {\rm GeV}$ (where we have used the energy loss rate
given in \eqref{fn_b}), and it becomes shorter as the energy
increases.  For such a scale, the change of the dark matter density is
not significant (except for the central region of the Galaxy).  In
particular, if the effect of the propagation is negligible, we can
omit the term proportional to $K(E)$ in the diffusion equation and, in
such a case, we obtain the following approximated formula for the
electron spectrum:
\begin{eqnarray}
  f_e^{\rm (local)} (E,\vec{x})
  = \frac{1}{\tau_{\rm DM}}
  \frac{\rho_{\rm DM}(\vec{x})}{m_{\rm DM}}
  \frac{ Y_e(>E)}{b(E,{\vec{x}})},
  \label{eq:f_e}
\end{eqnarray}
where
\begin{eqnarray}
  Y_e (>E) \equiv 
  \int_E^{\infty} dE^{\prime} \frac{dN_e}{dE^{\prime}}.
\end{eqnarray}
Thus, in such a case, the energy spectrum at the point $\vec{x}$ is
almost determined by the local dark matter density at the same point.  

\begin{figure}[t]
    \begin{center}
      \centerline{\epsfxsize=8.9 cm\epsfbox{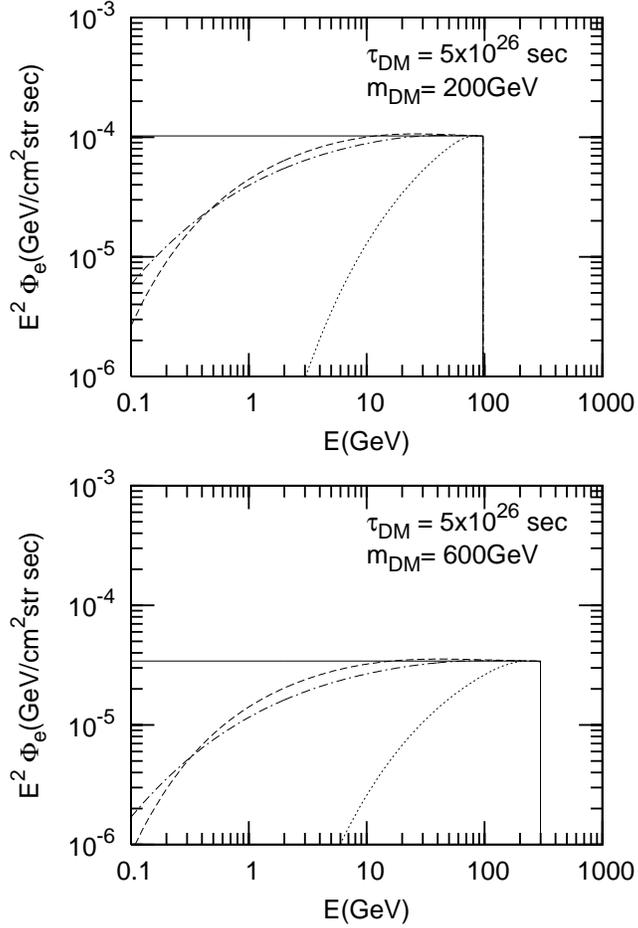}}
      \caption{Electron fluxes in MED (dashed), M1 (dot-dashed), and
        M2 (dotted) models, and that from $f_e^{\rm (local)}$ (solid),
        for the case that $Br(X\rightarrow e^+e^-)=1$.  Here, we take
        $m_{\rm DM}=200\ {\rm GeV}$ (top) and 600 GeV (bottom),
        $\tau_{\rm DM}=5 \times 10^{26}\ {\rm sec}$, $B=3 ~\mu$G, and
        $U_{\rm rad}=0.9 $ eV cm$^{-3}$.}
      \label{fig:positronfluxEE}
    \end{center}
    \vspace{-0.5cm}
\end{figure}

In Fig.\ \ref{fig:positronfluxEE}, we plot the electron flux $\Phi_e
\equiv \frac{c}{4\pi} f_e$ at the solar system (i.e., $r=r_\odot$ and
$z=0$).  Here, we consider the case that $Br(X\rightarrow e^+e^-)=1$,
taking $m_{\rm DM}=200\ {\rm GeV}$ and 600 GeV, $\tau_{\rm DM}=5
\times 10^{26}\ {\rm sec}$, and $B=3 ~\mu$G; then, the energy
distribution in Eq.\ \eqref{sourceterm} is
\begin{eqnarray}
  \frac{dN_{e}}{dE} = \delta (E - m_{\rm DM}/2).
  \label{dn/de(monochromatic)}
\end{eqnarray}
In addition, in depicting Fig.\ \ref{fig:positronfluxEE}, we use
$U_{\rm rad} \simeq 0.9$ eV cm$^{-3}$ for simplicity.  One can see
that the flux $f_e^{\rm (local)}$ well agrees with the predictions of
the MED and M1 models for $E \gtrsim 10$ GeV, which confirms our
earlier discussion.  On the other hand, one can observe a significant
difference between $f_e^{\rm (local)}$ and the flux in the M2 model.
This is due to the smallness of half-height of diffusion zone in the
M2 model, which is $1\ {\rm kpc}$; with such a small half-height, the
electron can escape from the diffusion zone.  Given the fact that the
MED model is the best fit to the cosmic-ray Boron to Carbon ratio
analysis, we adopt the MED model as our benchmark model of the
electron propagation.

\subsection{Synchrotron radiation: formalism}

Synchrotron radiation energy density per unit time and unit frequency
is expressed as
\begin{eqnarray}
  L_{\nu}(\vec{x})
  = \int dE \  {\cal P}(\nu,E) f_{e}(E,\vec{x}).
\end{eqnarray}
Here, ${\cal P}(\nu,E)$ is synchrotron radiation energy per unit time
and unit frequency from single electron with energy $E$, which is
given by \cite{Rybicki, Longair}
\begin{eqnarray}
  {\cal P}(\nu,E) 
  = \frac{1}{4\pi\epsilon_0} 
    \frac{\sqrt{3} e^3 B }{m_e c} F(\nu/\nu_c),
  \label{eq:sync_power}
\end{eqnarray}
where $\nu_c$ is critical frequency defined as
\begin{eqnarray}
  \nu_c \equiv \frac{3 eE^2B}{4\pi m^3_ec^4},
  \label{nu_c}
\end{eqnarray}
and 
\begin{eqnarray}
  F(x) \equiv x \int^{\infty}_{x} dy K_{5/3}(y),
\end{eqnarray}
with $K_{z}$ being the modified Bessel function of $z$-th
order.\footnote
{This formula, as well as energy loss rate given in Eq.\
  \eqref{eq:Psynch}, is for the case that the velocity of electron is
  perpendicular to the magnetic field.  Of course, such situation is
  not always satisfied in the Galaxy.  Therefore, considering the fact
  that the injected electron is isotropic and the magnetic field is
  entangled, we take the $B$ in the formula as mean magnetic flux
  density which the electron feels effectively.  }
The function $F(\nu/\nu_c)$ has a peak at $\nu\simeq 0.29\nu_c$.
Adopting the Galactic magnetic flux density of $B\sim3 \ \mu$G, we can
see that the synchrotron radiation in the observed frequency band of
the WMAP (i.e, $22-93$ GHz) is from the electron with the energy of $E
\sim 10-100$ GeV.\footnote
{In the previous works
  \cite{Gondolo:2000pn,Bertone:2001jv,Aloisio:2004hy,Regis:2008ij,
  Zhang:2008rs}, $F(x)$ is approximated as
  \begin{eqnarray*}
    F(x) \simeq \frac{8\pi}{9\sqrt{3}} \delta(x-0.29).
  \end{eqnarray*} 
  However, we found that such an approximation is not good in
  particular when $Y_e(>E)$ is suppressed for $E\gtrsim\sqrt{4 \pi
    m^3_e c^4 \nu/3 eB}$.}
For the electron in such an energy range, as we have seen, the
electron spectrum with the MED model, which is our benchmark model, is
well approximated by $f_e^{\rm (local)}$ given in Eq.\ (\ref{eq:f_e}).
Thus, we use $f_e^{\rm (local)}$ in our numerical calculation.

In order to calculate the observed radiation energy flux, we integrate
$L_{\nu}(\vec{x})$ along the line of sight (l.o.s.), whose direction is
parametrized by the parameters $\theta$ and $\phi$, where $\theta$ is
the angle between the direction to the Galactic center and that of the
line of sight, and $\phi$ is the rotating angle around the direction
to the Galactic center.  (The Galactic plane corresponds to $\phi=0$
and $\pi$.)  Then, the synchrotron radiation flux is given by
\begin{eqnarray}
  J_{\nu} (\theta, \phi)
  = \frac{1}{4 \pi} \int_{\rm l.o.s.} d \vec{l} L_{\nu}(\vec{l}).
\end{eqnarray}
We note here that we do not consider the synchrotron radiation from
the very central region of the Galaxy and that the effect of the self
absorption is unimportant \cite{Regis:2008ij}.  Using Eqs.\
(\ref{eq:f_e}) and (\ref{eq:sync_power}), we obtain the energy flux
from synchrotron radiation as
\begin{eqnarray} 
  J_{\nu}(\theta,\phi)
  =
  \frac{9\sqrt{3}}{32 \pi^2} \frac{1}{m_{\rm DM}\tau_{\rm DM}}
  \int_{\rm l.o.s.} d \vec{l} 
  \frac{\rho_{\rm DM}(\vec{l})}
  {1+r_{\rm IC/synch}(\vec{x})}
  \int dE \frac{Y_e(>E)}{\nu_c} F(\nu/\nu_c).
  \label{eq:flux}
\end{eqnarray}
Notice that, adopting the approximation of the constant magnetic flux
in the Galaxy, the line of sight and energy integrals factorize.

\section{Numerical Results}
\label{sec:results}
\setcounter{equation}{0} 

Now, we show the synchrotron radiation flux.  We numerically evaluate
the flux given in Eq.\ \eqref{eq:flux} for several dark matter models.
In the following, we show the results for $\phi=\frac{\pi}{2}$, i.e.,
$J_{\nu}(\theta,\pi/2)$.  The frequency of the radiation is taken to
be some of the WMAP frequency bands: $\nu=22$, $33$, and $61\ {\rm
  GHz}$.

\subsection{Leptonically decaying dark matter}

Let us start with the simplest case where the dark matter particle
dominantly decays as $X\rightarrow e^+e^-$.  This is the case if, for
example, the sneutrino field in supersymmetric model is the LSP, and
also if the $\hat{L}_i\hat{L}_1\hat{E}_1$-type $R$-parity violating
superpotential exists (where $\hat{L}_i$ and $\hat{E}_i$ are
superfields for $SU(2)_L$ doublet- and singlet-lepton superfields,
respectively) \cite{Ishiwata:2008cv}.  In this case, the energy
distribution in Eq.\ \eqref{sourceterm} is given by Eq.\
\eqref{dn/de(monochromatic)}.  As we have mentioned, with such a flux,
the PAMELA anomaly can be well explained if the lifetime of dark
matter is properly chosen.  (See Eq.\ \eqref{tau_ee}.)  In addition,
when $m_{\rm DM}$ is larger than $\sim 100\ {\rm GeV}$, the dark
matter contribution to the electron and positron fluxes for the energy
range of $100\ {\rm GeV}\lesssim E\leq \frac{1}{2}m_{\rm DM}$ are
approximately given by
\begin{eqnarray}
  E^2 \Phi_{e^\pm} \simeq 4.5 \times 10^{-5}\ {\rm GeV/cm^2\ sec\ str}
  \times 
  \left( \frac{\tau_{\rm DM}}{2.2\times 10^{26}\ {\rm sec}} \right)^{-1}
  \left( \frac{m_{\rm DM}}{1\ {\rm TeV}} \right)^{-1},
\end{eqnarray}
where the MED model of the propagation is used here.  (The flux for
$E\geq \frac{1}{2}m_{\rm DM}$ vanishes.) Then, if $m_{\rm DM}\sim 1\
{\rm TeV}$, a bump at $E\sim 500\ {\rm GeV}$ shows up in the
$(e^-+e^+)$ flux, taking the best-fit lifetime to explain the PAMELA
anomaly.  The behavior is consistent with the excesses of the
$(e^-+e^+)$ flux observed by the ATIC \cite{ATIC} and the PPB-BETS
\cite{Torii:2008xu} experiments.

\begin{figure}[t]
  \begin{center}
    \epsfxsize=0.9\textwidth\epsfbox{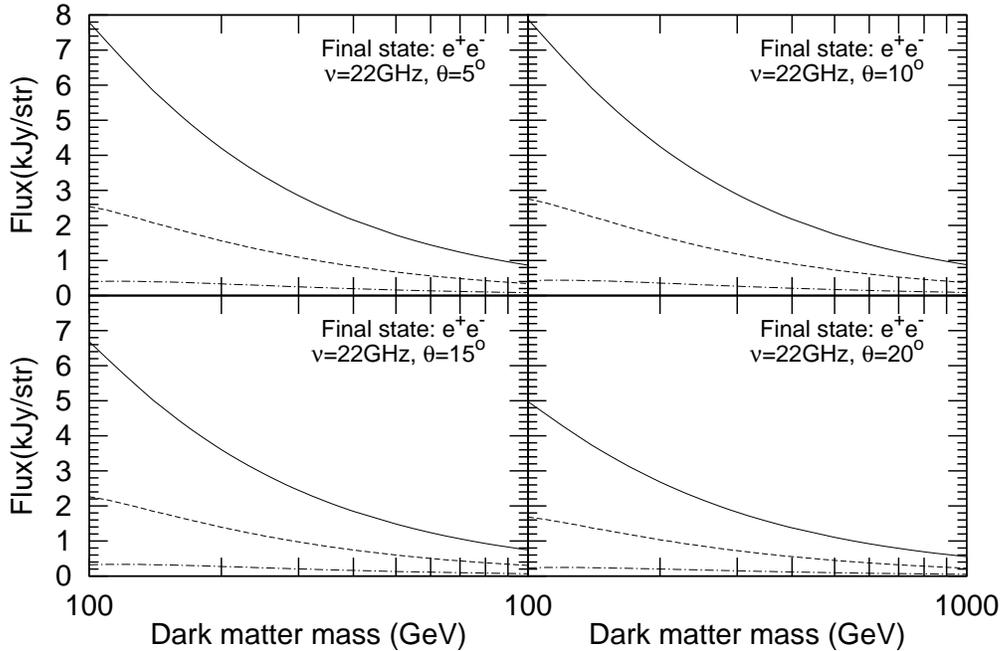}
    \caption{Synchrotron radiation fluxes at $\nu=22\ {\rm GHz}$ as
      functions of dark matter mass for angle $\theta = 5^{\circ}$,
      $10^{\circ}$, $15^{\circ}$, and $20^{\circ}$.  Here, the dark matter
      is assumed to decay only into $e^+ e^-$ pair.
      Here, we take $\tau_{3/2}=5 \times
      10^{26}$ sec, and show the cases of $B=$1, 3, 10 $\mu$G (from
      the bottom to the top) for each figure.}
    \label{fig:fluxEE_22}
  \end{center}
\end{figure}

\begin{figure}[t]
  \begin{center}
    \epsfxsize=0.9\textwidth\epsfbox{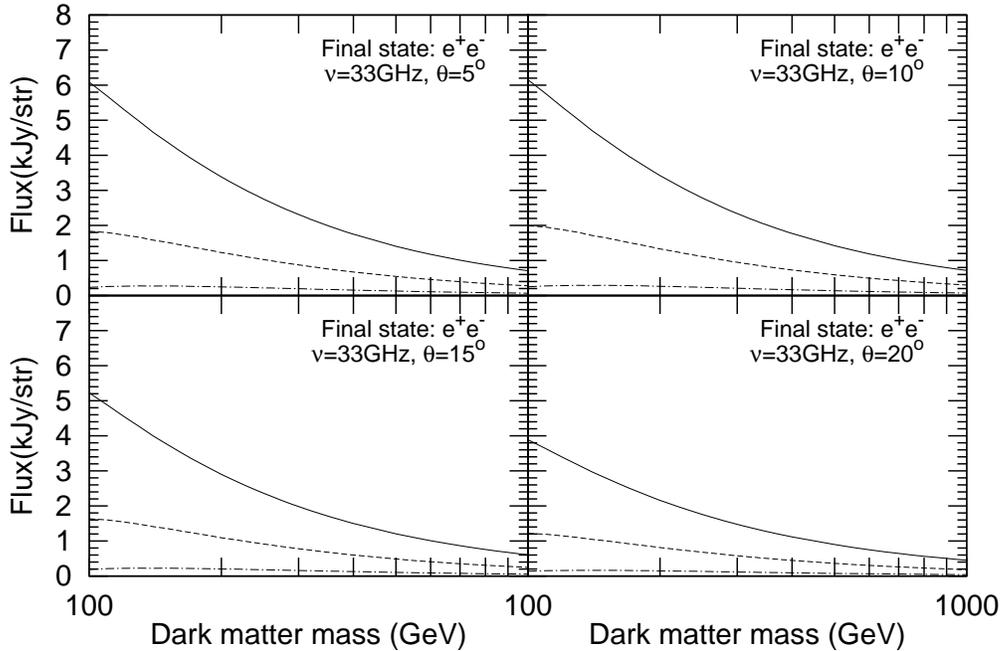}
    \caption{Same as Fig. \ref{fig:fluxEE_22}, except for $\nu=33\
      {\rm GHz}$.}
    \label{fig:fluxEE_33}
  \end{center}
\end{figure}

\begin{figure}[t]
  \begin{center}
    \epsfxsize=0.9\textwidth\epsfbox{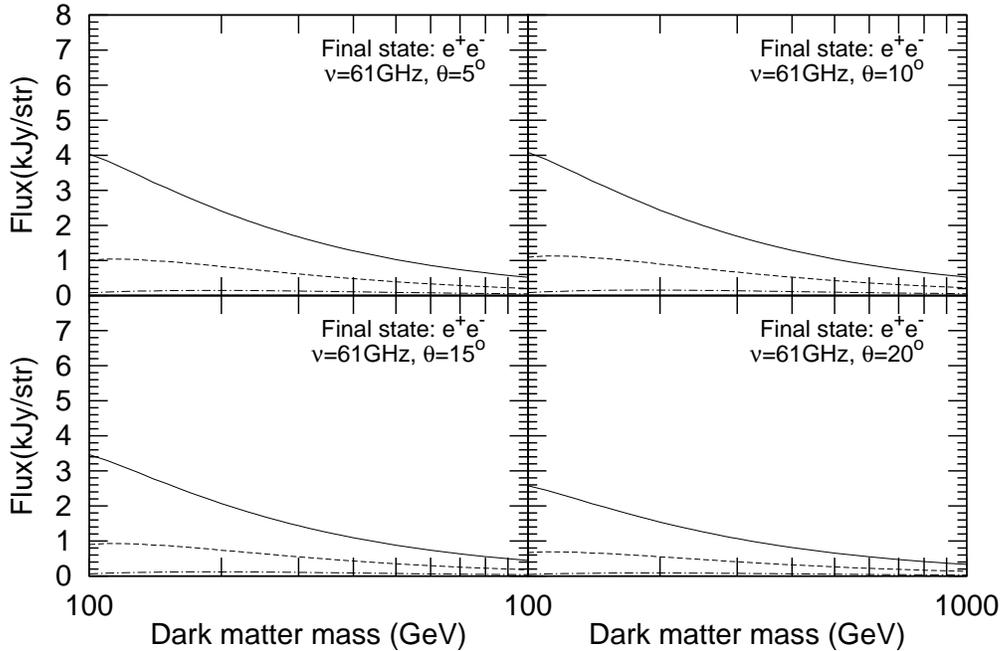}
    \caption{Same as Fig. \ref{fig:fluxEE_22}, except for $\nu=61\
      {\rm GHz}$.}
    \label{fig:fluxEE_61}
  \end{center}
\end{figure}

In Fig.\ \ref{fig:fluxEE_22}, we plot the fluxes for $\nu =22\ {\rm
  GHz}$ as functions of $m_{\rm DM}$, for $\tau_{\rm DM}=5\times
10^{26}$ sec, which is of the same order of the lifetime suggested
from the PAMELA anomaly.\footnote
{The synchrotron radiation flux is inversely proportional to
  $\tau_{\rm DM}$ (see Eq.(\ref{eq:flux})).  In order to obtain the
  flux for different value of $\tau_{\rm DM}$, one can simply rescale
  the flux given in the figure.}
The angle $\theta$ is taken to be $5^{\circ}$, $10^{\circ}$,
$15^{\circ}$, and $20^{\circ}$ (which give the minimum distance from
the Galactic center to the line of sight of $0.7$, $1.5$, $2.3$, and
$3.1\ {\rm kpc}$, respectively), and $B=1$, $3$, and $10\ \mu{\rm G}$.
We also show the fluxes for $\nu =33$ and $61\ {\rm GHz}$ in Figs.\
\ref{fig:fluxEE_33} and \ref{fig:fluxEE_61}, respectively.  

As one can see, the synchrotron radiation flux from the decay is $O(1\
{\rm kJy/str})$ or smaller in wide parameter region $m_{\rm DM} \sim
100\ {\rm GeV}-1\ {\rm TeV}$.  The flux $J_{\nu}$ decreases as the
frequency increases.  In addition, $J_{\nu}$ decreases as $B$
decreases; this is because, as the magnetic field becomes weaker, the
$e^\pm$s lose more energy by inverse Compton process, resulting in a
suppression of the synchrotron radiation flux.

Radiation flux from Galactic center region has been observed by the
WMAP for frequency bands of 22, 33, 41, 61, and 93 GHz
\cite{Dobler:2007wv, Gold:2008kp}.  Since then, intensive analysis has
been performed to understand the origins of the radiation flux.  (For
recent studies, see \cite{Dobler:2007wv, Hooper:2007kb, Hooper:2008zg,
  Gold:2008kp}.)  Most of the radiation flux is expected to be from
astrophysical origins, such as thermal dust, spinning dust, ionized
gas, and synchrotron radiation, which have been studied by the use of
other survey data
\cite{Finkbeiner:1999aq,Finkbeiner:2003im,Finkbeiner:2003yt}.  With
the three-year data, the WMAP collaboration claimed that the flux
intensity can be explained by the known astrophysical origins
\cite{Hinshaw:2006ia}.  On the contrary, Refs.\ \cite{Hooper:2007kb,
  Hooper:2008zg, Dobler:2007wv} also studied the WMAP three-year data,
and claimed that there exists a remnant flux from unknown origin which
might be non-astrophysical; the remnant flux is called the ``WMAP
Haze''.\footnote
{In \cite{Hooper:2007kb, Hooper:2008zg}, as a possible explanation
  for the WMAP Haze, annihilation of dark matter is proposed.  If
  energetic charged particles are emitted via the dark matter
  annihilation, they become another source of the synchrotron
  radiation.  According to the study, the synchrotron radiation flux
  turns out to be comparable to the WMAP Haze with the annihilation
  cross section $\langle\sigma v\rangle\sim O(10^{-26}\ {\rm cm}^3{\rm
    sec}^{-1}$), which is consistent magnitude to explain dark matter
  abundance.}
However, no clear indication of the WMAP Haze from unknown source was
reported by the WMAP collaboration after five-year data
\cite{Gold:2008kp}.

The existence of the WMAP Haze seems still controversial, and the
detailed studies of the WMAP Haze using the data is beyond the scope of
our study.  Here, we adopt the flux of the WMAP Haze suggested in
\cite{Hooper:2007kb, Hooper:2008zg} as a reference value.  Our numerical
calculation shows that the predicted flux, which is $O(1\ {\rm
kJy/str})$, is comparable to the flux of the WMAP Haze given in
\cite{Hooper:2007kb, Hooper:2008zg}.  As we mentioned, since the the
existence of the exotic radiation flux of this size is controversial, it
is difficult to confirm or exclude the present scenario without better
understandings of the sources of Galactic foreground emission.

\subsection{Gravitino dark matter}

Next, let us consider the case that the gravitino (which we denote
$\psi_\mu$) is the LSP and is dark matter.  Even if the gravitino is
the LSP, it becomes unstable if the $R$-parity is violated.  In
particular, if the effect of the $R$-parity violation is small enough,
the lifetime of the gravitino becomes longer than the present age of
the universe and it is a viable candidate for dark matter
\cite{Takayama:2000uz}.  Production of the gravitino may be due to the
thermal scattering \cite{Moroi:1993mb}, the decay of the lightest
superparticle in the minimal supersymmetric standard model sector
\cite{Feng:2003xh, Feng:2003uy, Ellis:2003dn, Roszkowski:2004jd,
  Ishiwata:2007bt}, and the decay of the inflaton field
\cite{Endo:2007ih}.  We do not specify any particular scenario, but
just assume that the density parameter of the gravitino has somehow
become consistent with the dark matter density.

In this study, we consider the following $R$-parity violating
operators:
\begin{eqnarray}
  {\cal L}_{\rm RPV} 
  = B_i \tilde{L}_i H_u + m^2_{\tilde{L}_i H_d} \tilde{L}_i H^*_d 
  + {\rm h.c.},
  \label{L_RPV}
\end{eqnarray}
where $\tilde{L}_i$ is left-handed slepton doublet in $i$-th
generation, while $H_u$ and $H_d$ are up- and down-type Higgs boson
doublets, respectively.  (We work in the basis where the bi-linear
$R$-parity violating superpotential vanishes, which is realized by an
appropriate redefinition of the Higgs and lepton-doublet chiral
superfields.)  With the above-mentioned $R$-parity violating
operators, the anomalous excesses of the positron
\cite{Barwick:1997ig, Adriani:2008zr} and $\gamma$-ray
\cite{Sreekumar:1997un} in the cosmic ray can be simultaneously
explained when the parameters $B_i$ and $m^2_{\tilde{L}_i H_d}$ are
properly chosen.\footnote
{It is often the case that, when the decay of dark matter produces
hadronic objects, like $q\bar{q}$ pair, the cosmic-ray anti-proton
flux becomes too large to be consistent with observations.  However,
the anti-proton flux is sensitive to the model of cosmic-ray
propagation \cite{Ibarra:2008qg}.  In addition, the ratio of the
number of the $q\bar{q}$ pair to that of the positron is $\sim 6$ in
the present scenario, which is about $4$ times larger than the ratio
for the scenario proposed in \cite{Chen:2008md} where the heavy gauge
boson for an exotic $U(1)$ gauge symmetry, which has a small kinetic
mixing with the $U(1)_{B-L}$ gauge boson, is dark matter.  The
cosmic-ray anti-proton flux in such a scenario is order of magnitude
smaller than the observed anti-proton flux.  Thus, we expect that the
constraint from the anti-proton flux does not exclude the present
scenario.  More detail of this subject will be discussed elsewhere
\cite{IMM_WorkInProgress}.}

The flavor of the final-state lepton depends on the sizes of the
$R$-parity violating coupling constants.  In this analysis, for
simplicity, we assume that the gravitino decays only into leptons in
one of the three generations.  Then, the gravitino decays as
$\psi_{\mu}\rightarrow l_f^- W^+$, $\nu_f Z$, $\nu_f h$, and
$\nu_f\gamma$ (where $f=1$, $2$, or $3$ specifies the generation of
the final-state lepton).  In particular, the process
$\psi_{\mu}\rightarrow l_f^\pm W^\mp$ has the largest branching ratio
(when the process is kinematically allowed).  Notably, from this
process, the charged lepton $l_f^\pm$ is directly produced, which
results in an energetic electron or positron.  If $l_f^\pm=e^\pm$,
such an $e^\pm$ has energy of $(m_{3/2}^2-m_W^2)/2m_{3/2}$, where
$m_{3/2}$ is the gravitino mass.  In addition, the decays of the weak-
and Higgs-bosons also produce $e^\pm$.  As we mentioned in Section\
\ref{sec:scenario}, the predicted positron fraction well agrees with
the PAMELA result if the lifetime of the gravitino is given by Eq.\
\eqref{tau_gravitino}.  In addition, in such a case, when $m_{3/2}$ is
large enough, the electron and positron fluxes for the energy range of
$100\ {\rm GeV}\lesssim E\leq \frac{1}{2}m_{\rm DM}$ are approximately
given by
\begin{eqnarray}
  E^2 \Phi_{e^\pm} \simeq 3.0 \times 10^{-5}\ {\rm GeV/cm^2\ sec\ str}
  \times 
  \left( \frac{\tau_{3/2}}{8.5\times 10^{25}\ {\rm sec}} \right)^{-1}
  \left( \frac{m_{3/2}}{1\ {\rm TeV}} \right)^{-1},
\end{eqnarray}
where $\tau_{3/2}$ is the lifetime of the gravitino.  Then, as in the
previous case, there exists a bump in the $(e^-+e^+)$ flux at $E\sim
500\ {\rm GeV}$ if we take $m_{3/2}\sim 1\ {\rm TeV}$ (and the PAMELA
best-fit value of the lifetime).  This may be the origin of the excess
of the $(e^-+e^+)$ flux recently observed by the ATIC \cite{ATIC} and
the PPB-BETS \cite{Torii:2008xu} experiments.  (For detail, see
\cite{IMM_WorkInProgress}.)  On the contrary, if $l_f^\pm=\mu^\pm$ and
$\tau^\pm$, $e^\pm$ from the decay of $l_f^\pm$ becomes
non-monochromatic.

The electron and positron produced by the gravitino decay become the
sources of the synchrotron radiation.  In our numerical calculation, we
include all the relevant interaction terms to calculate the partial
decay rates; for detail, see \cite{Ishiwata:2008cu}.\footnote
{There are typos in Eqs.\ (3.18) and (3.19) of \cite{Ishiwata:2008cu}.
  The factors in front of the terms proportional to the function $G$
  should be $\frac{2}{3}$, not $\frac{3}{2}$.  We thank L. Covi for
  pointing out these typos.}
Then, for a precise calculation of the energy spectrum, we use the
PYTHIA package \cite{Sjostrand:2006za} to calculate $dN_{e}/dE$.

\begin{figure}[t]
  \begin{center}
    \epsfxsize=0.9\textwidth\epsfbox{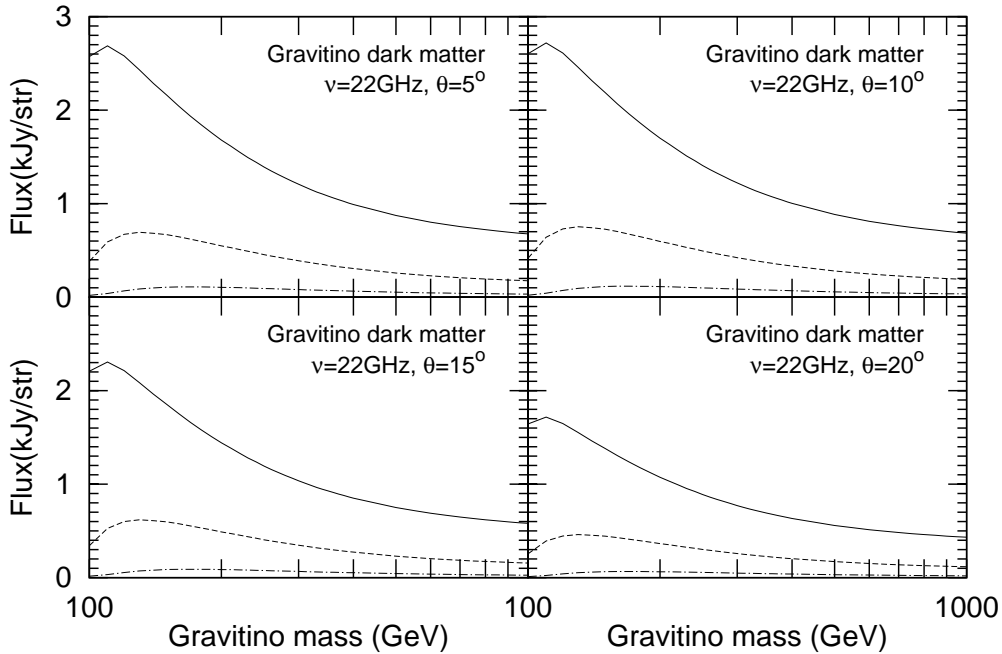}
    \caption{Synchrotron radiation fluxes at $\nu=22\ {\rm GHz}$ as
      functions of gravitino mass for angle $\theta = 5^{\circ}$,
      $10^{\circ}$, $15^{\circ}$, and $20^{\circ}$.  The final-state
      lepton in the gravitino decay is in the first generation.  Here,
      we take $\tau_{3/2}=5 \times 10^{26}$ sec, and show the cases of
      $B=$1, 3, 10 $\mu$G (from the bottom to the top) for each
      figure.}
    \label{fig:fluxgr2e_22}
  \end{center}
\end{figure}

First, we show the results for the case where $f=1$ (i.e., $l_f^\pm
=e^\pm$), for which $J_{\nu}$ is maximized for the gravitino dark
matter case.  In Fig.\ \ref{fig:fluxgr2e_22}, we plot the fluxes for
$\nu =22\ {\rm GHz}$ as functions of the gravitino mass, taking
$\tau_{3/2}=5\times 10^{26}$ sec.  The angle $\theta$ is taken to be
$5^{\circ}$, $10^{\circ}$, $15^{\circ}$, and $20^{\circ}$, and $B=1$,
$3$, and $10\ \mu{\rm G}$.  One can see that the flux is maximized
when the dark matter mass is slightly larger than $100\ {\rm GeV}$.
This is due to the fact that, when $m_{\rm DM} \sim 100\ {\rm GeV}$,
$\nu_c$ for $e^\pm$ from the decay of $X$ defined in Eq.\ \eqref{nu_c}
becomes comparable to $\nu\sim O(10\ {\rm GHz})$.

\begin{figure}[t]
  \begin{center}
    \epsfxsize=0.9\textwidth\epsfbox{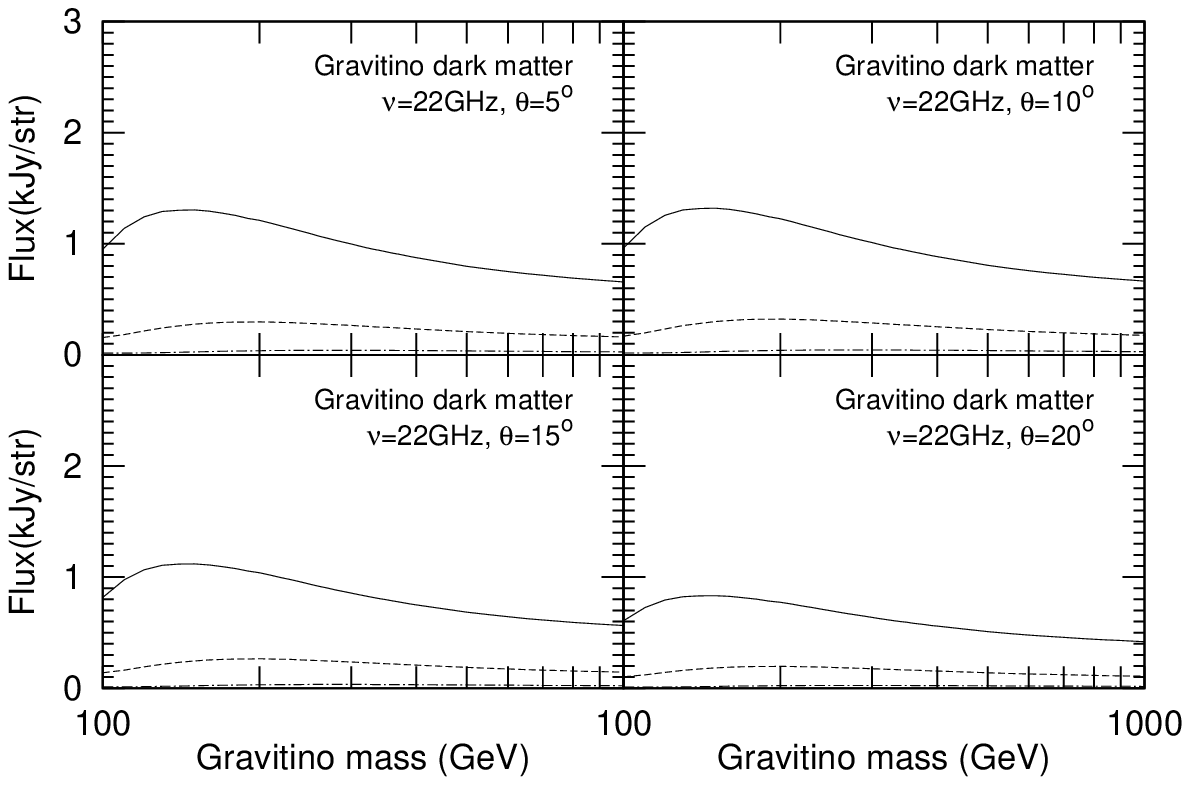}
    \caption{Same as Fig.\ \ref{fig:fluxgr2e_22}, except for the case
      that the final-state lepton in the gravitino decay is in the
      second generation.}
    \label{fig:fluxgr2mu_22}
    \end{center}
    \vspace{-0.5cm}
\end{figure}

The second case is that the primary lepton produced by the gravitino
decay is in the second generation.  The results are shown in Fig.\
\ref{fig:fluxgr2mu_22}.  The synchrotron radiation flux becomes
smaller in this case, compared to the case of $f=1$.  This is because
the $e^\pm$ produced by the decay of $\mu^\pm$ is less energetic than
that directly produced by the process $\psi_{\mu} \rightarrow e^\pm
W^\mp$.  We have also studied the case that $l_f^\pm =\tau^\pm$.  As
expected, the radiation flux becomes more suppressed in such a case.

For the gravitino dark matter case, the synchrotron radiation flux is
again of the order of $\sim 1\ {\rm kJy/str}$ or smaller.  As we have
mentioned, we expect that the synchrotron radiation flux of this size
is not excluded by the results of the presently available
observations.  The flux in the gravitino dark matter case is smaller
than that for the dark matter which decays into $e^+ e^-$ pair (if the
lifetime is fixed).  This is due to the fact that the total amount of
energy carried away by $e^\pm$ is smaller in the case of the gravitino
dark matter.

\section{Summary}
\label{sec:summary}
\setcounter{equation}{0}

In this paper, we have discussed the synchrotron radiation flux from
the Galactic center in unstable dark matter scenario.  Motivated by
the recently reported PAMELA positron excess, we consider unstable
dark matter whose decay produces energetic electron and positron; if
the lifetime of dark matter is $O(10^{26-27}\ {\rm sec})$, the
observed positron flux can be well explained in such a scenario.  Once
the energetic $e^\pm$ is produced in the Galaxy, it becomes the source
of the synchrotron radiation.  As discussed in Section
\ref{sec:synchotron}, in our study, we have used a formalism with
several improvements to calculate the synchrotron flux.  Then, we have
numerically calculated the flux for models which can well explain the
PAMELA anomaly.

Assuming $\tau_{\rm DM}\sim O(10^{26}\ {\rm sec})$ to explain the
PAMELA anomaly, we found that the synchrotron radiation flux from the
dark matter decay is expected to be $J_\nu\sim O(1\ {\rm kJy/str})$
(or smaller).  As we have mentioned, the existence of the exotic
radiation flux of this size is controversial.  However, it should be
also noted that the synchrotron radiation flux is inversely
proportional to the lifetime of dark matter.  In order to explain the
PAMELA anomaly, the smaller value of $\tau_{\rm DM}$ is preferred as
$m_{\rm DM}$ becomes lager.  (See Eqs.\ \eqref{tau_ee} and
\eqref{tau_gravitino}.)  Thus, if we consider too large $m_{\rm DM}$,
it may become difficult to explain the PAMELA anomaly in the decaying
dark matter scenario once we take into account the WMAP observation of
the radiation from the central region of our Galaxy.

In any case, without better understandings of the sources of Galactic
foreground emission, it is difficult to confirm or exclude the
scenario of explaining the PAMELA anomaly in the decaying dark matter
scenario (with $m_{\rm DM}\sim O(100\ {\rm GeV})$).  A more detailed
study of the Galactic emission may provide a significant test of the
unstable dark matter scenario.  Once the existence of the WMAP Haze
will be somehow confirmed, decaying dark matter may be a viable
candidate of its origin.

\noindent {\it Acknowledgements:} This work was supported in part by
Research Fellowships of the Japan Society for the Promotion of Science
for Young Scientists (K.I.), and by the Grant-in-Aid for Scientific
Research from the Ministry of Education, Science, Sports, and Culture
of Japan, No.\ 19540255 (T.M.).

\end{document}